# Accelerating Physics-Informed Neural Network based 1-D arc simulation by meta learning

Linlin Zhong [1,a], Bingyu Wu [1], Yifan Wang [1]

[1] *School of Electrical Engineering, Southeast University, No 2 Sipailou, Nanjing, Jiangsu Province 210096, P. R. China*

Physics-Informed Neural Networks (PINNs) have a wide range of applications as an alternative to traditional numerical methods in plasma simulation. However, in some specific cases of PINN-based modeling, a well-trained PINN may require tens of thousands of optimizing iterations during training stage for complex modeling and huge neural networks, which is sometimes very time-consuming. In this work, we propose a meta-learning method, namely Meta-PINN, to reduce the training time of PINN-based 1-D arc simulation. In Meta-PINN, the meta network is first trained by a two-loop optimization on various training tasks of plasma modeling, and then used to initialize the PINN-based network for new tasks. We demonstrate the power of Meta-PINN by four cases corresponding to 1-D arc models at different boundary temperatures, arc radii, arc pressures, and gas mixtures. We found that a well-trained meta network can produce good initial weights for PINN-based arc models even at conditions slightly outside of training range. The speed-up in terms of relative $L^2$ error by Meta-PINN ranges from 1.1× to 6.9× in the cases we studied. The results indicate that Meta-PINN is an effective method for accelerating the PINN-based 1-D arc simulation.

## 1. Introduction

As the fourth state of matter, plasmas have a wide range of applications in laboratories and industry. To understand and control plasma behavior in various plasma assisted applications, many computational plasma physics models have been developed in the past few decades. For instance, the particle-in-cell (PIC) simulation [1] and the plasma fluid simulation (e.g., Magnetohydrodynamics (MHD) modeling [2]) are two typical representatives, which can describe a plasma from a microscopic and macroscopic view, respectively. In whichever plasma modeling, the nature of simulation is solving plasma governing equations, i.e. some partial differential equations (PDEs) mathematically describing plasma behavior. The typical procedure of plasma simulation includes at least the sophisticated meshing on the defined computational domain, the discretization of plasma PDEs on the meshes, and the numerical solving of discretized PDEs. Due to the irregular geometry and complex coupling relationship of multiphysics in real-world applications, the classical plasma modeling sometimes encounters difficulties in meshing, discretizing, and numerical solving, leading to poor convergence and even divergence of plasma simulation. As a result, many efforts are being directed on alternative techniques to classical plasma modeling.

Recently, the Physics-Informed Neural Networks (PINNs) proposed by Raissi, Karniadakis and their cooperators [3, 4] have drawn much attention in the field of scientific computation. PINNs incorporate the physics constraints described by PDEs in the loss function of neural networks and approximate the solution by minimizing the loss value. Compared with traditional numerical methods, such as the finite difference method (FDM), the finite element method (FEM), and the finite volume method (FVM), the PINN-based methods are mesh-free and require no handcrafted discretization of PDEs. This makes PINNs easy to use and have greate potential in scientific computing. Also because of this, PINNs have been more and more applied in solving physics, mechanics, chemistry, and biology problems [4-8]. In the field of plasma physics, PINNs are also increasingly gaining attention and applied in, for example, solving the Boltzmann equation of weakly ionized plasmas [9] and inversing the parameters of inhomogenous magnetized plasmas [10]. In our previous works [11, 12], the PINN-based simulation frameworks, namely the Coefficient-Subnet Physics-Informed Neural Network (CS-PINN) and Runge–Kutta Physics-Informed Neural Network (RK-PINN), were proposed for the kinetic and fluid modeling of low-temperature plasmas. As a meshless method, PINNs allow easy implementation and highly adaptive to various plasma simulation cases. However, as a data-driven method, PINNs have a large number of parameters (i.e. the weights of neural networks) to be learned from data, which is usually performed by iterative training in terms of gradient descent. In some specific cases of PINN-based modeling, the training of PINNs is sometimes very time-consuming especially for large and deep neural networks. This could weaken the advantages of PINNs over traditional numerical methods. Even worse, PINNs are task specific, which means a well trained PINN for solving a specific problem cannot be used directly for a new problem despite a small change, e.g., a slight variation of a coefficient in a PDE. This is because the ability to learn new concepts quickly is limited for neural networks [13].

[a] mathboylinlin@gmail.com, linlin@seu.edu.cn

With the development of machine learning, meta learning has been suggested as an approach to address the aforementioned issue by enabling the neural networks to learn how to learn. The key idea of meta learning is to train the neutral networks on various tasks instead of just on various data of a single task. This means the learning ability of neural networks can be generalized from task to task. Similarly, training PINNs on a new task can also be accelerated by meta learning. Psaros et al. [14] have demonstrated that the performance of PINNs can be improved significantly by meta-learning loss functions. Since the ultimate goal of training is to determine the optimum weights of a neural network, why not learn the weights directly through meta learning? This is exactly what the meta-learning initialization does. Inspired by this, we propose a meta initialization method in this work for accelerating PINN-based plasma simulation and a 1-D arc model is used an example to validate this method. We anticipate that this method would have the ability of improving the learning efficiency of PINNs on different tasks of plasma simulation.

The remaining part of the paper is organized as follows. In Section 2, the framework of PINN-based plasma simulation is first introduced briefly and the description of a meta-learning framework, namely Meta-PINN, is succeeded for accelerating PINN-based plasma simulation. In Section 3, we demonstrate the performance of Meta-PINN by four cases including solving 1-D arc models at different boundary temperatures, arc radii, arc pressures, and gas mixtures, respectively. Lastly a summary is concluded in Section 4.

## 2. Meta-learning framework for accelerating plasma simulation

### 2.1. Physics-Informed Neural Network (PINN) based plasma simulation

Since our prior studies [11, 12] have explored the frameworks of PINN-based plasma simulation, we just present a quick overview here. Let's consider the following general plasma equation defined on spatiotemporal domain Ω [11].

$$f(t, \mathbf{x}, \partial_t u, \partial_{\mathbf{x}} u, \cdots, \boldsymbol{\lambda}) = 0 \tag{1}$$

With boundary and initial conditions defiend on ∂Ω

$$\mathcal{B}(u, t, \mathbf{x}) = 0, \quad \mathcal{I}(u, \mathbf{x}) = 0 \tag{2}$$

Where $\mathbf{x} \in \mathbb{R}^d$ and $t$ are the spatial and temporal coordinates, respectively; $u(t, \mathbf{x})$ is the solution, and $\boldsymbol{\lambda} = [\lambda_1, \lambda_2, \cdots]$ are the coefficients in the plasma equation, which are usually thermodynamic and transport properties of a plasma, such as mass density, specific heat, electrical conductivity, and thermal conductivity.

In the framework of PINNs, the solution $u(t, \mathbf{x})$ of the equation (1) is approximated by constructing a neural network with the trainable parameters θ. The network receives $(t, \mathbf{x})$ as input and produces $\hat{u}(t, \mathbf{x})$ as output. Using the technique of automatic differentiation, the plasma equations including initial and boundary conditions can be easily expressed by the neural network. PINNs aim to find a network with optimum weights that satisfy the governing equations as well as possible. This is usually achieved by minimizing the loss function $\mathcal{L}_{PINN}(\theta)$ defined as follows [11].

$$\mathcal{L}_{PINN}(\theta) = \frac{\omega_f}{N_f} \sum_{i=1}^{N_f} \mathcal{F}\left(f(t_i, \mathbf{x}_i, \partial_t \hat{u}_i, \partial_{\mathbf{x}} \hat{u}_i, \cdots, \boldsymbol{\lambda})\right) + \frac{\omega_{\mathcal{B}}}{N_{\mathcal{B}}} \sum_{i=1}^{N_{\mathcal{B}}} \mathcal{F}\left(\mathcal{B}(\hat{u}_i, t_i, \mathbf{x}_i)\right) + \frac{\omega_{\mathcal{I}}}{N_{\mathcal{I}}} \sum_{i=1}^{N_{\mathcal{I}}} \mathcal{F}\left(\mathcal{I}(\hat{u}_i, \mathbf{x}_i)\right) \tag{3}$$

Where $\omega_f$, $\omega_{\mathcal{B}}$, and $\omega_{\mathcal{I}}$ are the weighting factors regulating the loss values relating to governing equations, boundary conditions, and initial conditions, respectively; $N_f$, $N_{\mathcal{B}}$, and $N_{\mathcal{I}}$ are accordingly the numbers of the scattered points for different loss terms; $\mathcal{F}(\bullet)$ is the function that measures the residuals of equations.

Considering that various plasma properties are strongly dependent on plasma compositions, temperatures, and pressures, we proposed CS-PINN in the previous work [11], which uses a subnet (i.e. a neural network or an interpolation function) to express the coefficients $\boldsymbol{\lambda}$ in the plasma equations.

## 2.2. Meta learning for PINN-based plasma simulation

As introduced in Section 2.1, the PINN-based plasma simulation uses a neural network as the surrogate of plasma equations. Typically, the network is initialized randomly and then trained by a gradient descent algorithm until the loss value is small enough. The problem is that such training of PINNs starting from random initialization needs tens of thousands of iterations to find the optimum weights in the vast parameter space. This is extremely time-consuming for the PINNs with huge neural networks and large amounts of training data.

To improve the efficiency in training PINNs for plasma simulation, we follow the work by Finn et al. [15] and propose a meta-learning method as an effective approach to initializing PINNs, which is called Meta-PINN in this work. The core idea is as illustrated in Figure 1(a). We first train a meta network on various training tasks of plasma modeling. For a given new task (i.e. a testing task), the weights of PINNs are initialized by the meta network instead of random initialization before starting the normal training. As shown in Figure 1(a), there are two types of networks, i.e. the PINN network and the meta network. The former is the neural network as we described in Section 2.1. It is the surrogate of a plasma model which can be any form of PINNs, e.g. CS-PINN or other variants, and is trained on the scattered data of the computational domain by minimizing the loss function $\mathcal{L}_{PINN}(\theta)$. The latter is the meta neural network which learns meta knowledge from different training tasks by minimizing the meta loss $\mathcal{L}_{meta}(\theta)$.

$$\mathcal{L}_{meta}(\theta) = \frac{1}{N_{task}} \sum_{i=1}^{N_{task}} \mathcal{L}_{PINN}(\theta_i) \qquad (4)$$

Where $N_{task}$ is the number of training tasks which are sampled from a task distribution space. In plasma modeling, the training tasks can be composed of plasma models with different parameters, e.g., the plasma equations with different coefficients.

According to the paradigm of meta learning, the PINN network and the meta network should be trained on different data to ensure the generalization. Following the work by Ni et al. [16], we split the scattered training set by support data and query data. The PINN and meta networks are trained on support data and query data respectively. As described in Figure 1(b) and Algorithm 1, the whole training is a two-loop optimization. The PINN network is optimized in the inner loop by the stochastic gradient descent (SGD) algorithm. After several iterations of the inner loop, the meta loss is calculated based on query data and then optimized in the outer loop also by SGD algorithm. With the decrease of meta loss, the meta network is learning weights as meta knowledge. After the meta training, the meta network would have the power of yielding good weights as initialization for a PINN network in a testing task which is different from the training tasks. We will demonstrate in Section 3 that the initialized weights by the meta network are much better than the weights initialized by the widely used technique i.e. the Xavier initialization [17] which has been verified to performs better than random initialization. This will speed-up the training of PINNs for plasma simulation. It is notable that some neural operator (NO)-based models, such as Fourier neural operator (FNO) [18] and deep operator network (DeepONet) [19], can also deal with the cases studied in this work in the way of operator learning. We provide an alternative approach (i.e. the meta-learning method in this work) to solving parametric arc models. The performance comparison between NOs and meta-learning method for plasma simulation still needs further investigation in future work.

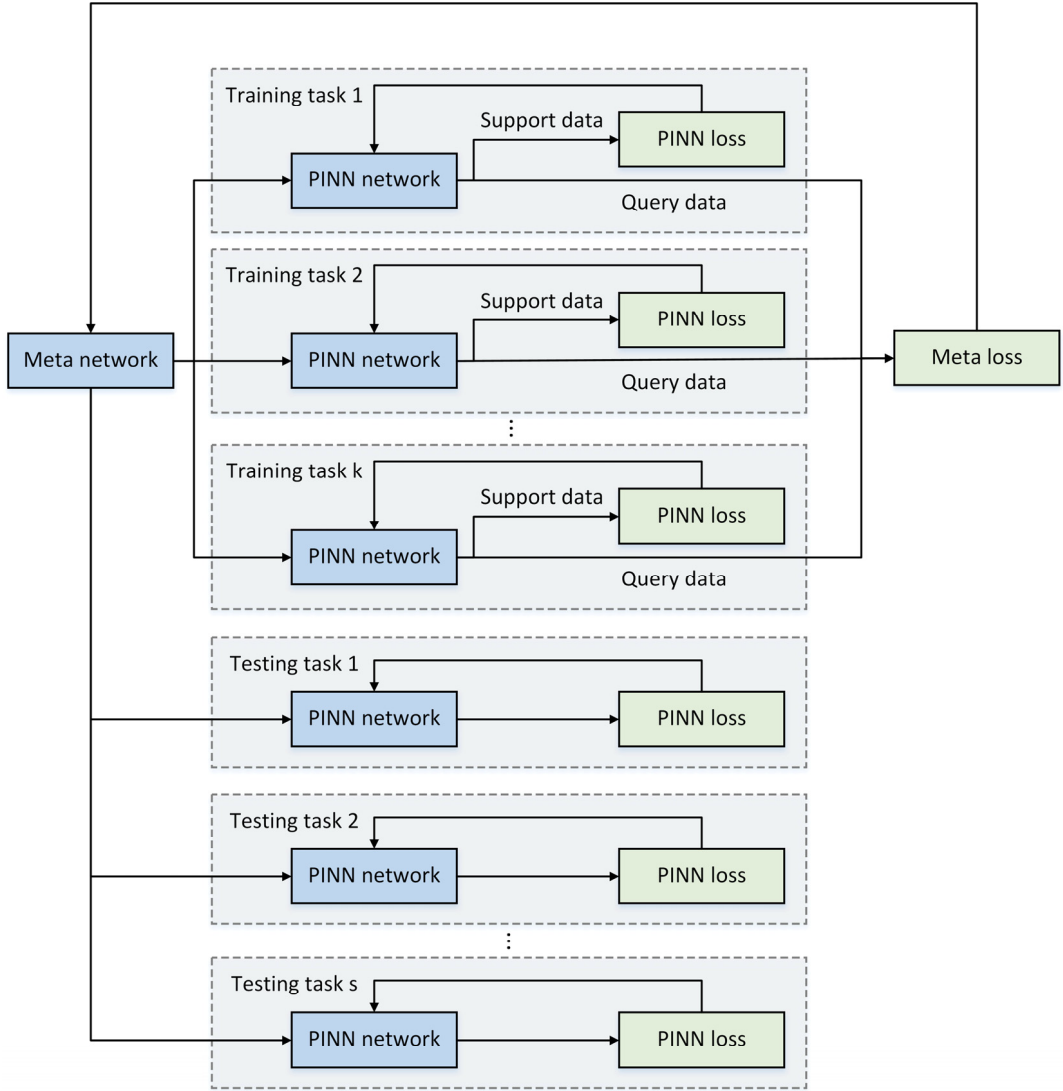

(a) Training and testing of meta network

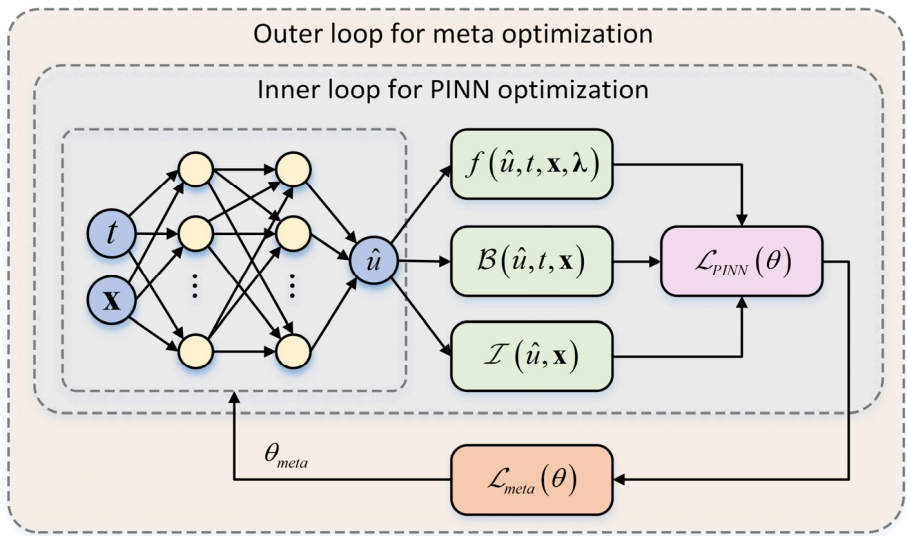

(b) Two-loop optimization: Inner loop for PINN optimization and outer loop for meta optimization

**Figure 1**. Diagram of Meta-PINN for plasma simulation

---

**Algorithm 1:** Meta-PINN for plasma simulation

**Input:** task distribution $P$, number of training tasks $N_{task}$, number of inner iterations $N_{inner}$, number of outer iterations $N_{outer}$, inner learning rate $\eta_{inner}$, outer learning rate $\eta_{outer}$

**Output:** weights of meta network $\theta_{meta}$

1: initialize $\theta_{meta}$ randomly or by a specific technique, e.g. Xavier initialization
2: initialize the meta loss $\mathcal{L}_{meta}(\theta_{meta}) = 0$
3: **for** $i = 1, \ldots, N_{outer}$
4:   **for** $j = 1, \ldots, N_{task}$
5:     sample training task $j$ from $P$
6:     sample support data for task $j$
7:     initialize the PINN network $\theta_j$ for task $j$ by $\theta_{meta}$
8:     **for** $k = 1, \ldots, N_{inner}$
9:       calculate the PINN loss $\mathcal{L}_{PINN}(\theta_j)$
10:       update $\theta_j$ by minimizing $\mathcal{L}_{PINN}(\theta_j)$ with learning rate $\eta_{inner}$
11:     **end for**
12:     sample query data for task $j$
13:     calculate the meta loss $\mathcal{L}_{meta}(\theta_j)$ on query data of task $j$
14:     update the meta loss $\mathcal{L}_{meta}(\theta_{meta}) += \mathcal{L}_{meta}(\theta_j)$
15:   **end for**
16:   update $\theta_{meta}$ by minimizing $\mathcal{L}_{meta}(\theta_{meta})$ with learning rate $\eta_{outer}$
17: **end for**
18: **return** $\theta_{meta}$

---

## 3. Case study of 1-D arc model

### 3.1. Model description

To validate the performance of Meta-PINN, we conduct a case study of 1-D arc model as an example. This is a stationary arc in a cylindrical column mathematically described by the Elenbass-Heller equation [12].

$$\frac{1}{r}\frac{\partial}{\partial r}\left(r\kappa\frac{\partial T}{\partial r}\right) + \sigma\frac{I^2}{g^2} - E_{rad} = 0 \tag{5}$$

$$g = \int_0^R 2\pi r\sigma dr \tag{6}$$

Where $r$ is the radial coordinate, $T$ the arc temperature, $\sigma$ the electrical conductivity, $\kappa$ the thermal conductivity,

*I* the current, *g* the arc conductance, $E_{rad}$ the radiation energy, and *R* the arc radius. The zero temperature gradient and constant temperature are used as the boundary conditions at *r* = 0 and *r* = *R* respectively.

Using the framework of CS-PINN [11], we construct a feed-forward neural network with 6 hidden layers and 50 neurons in each layer as the surrogate of the arc model. The spline functions are used as subnet to approximate plasma properties in the model. In this work, $SF_6$ and $SF_6$-$N_2$ arc plasmas are considered as the examples. All the transport and radiation coefficients for modeling $SF_6$ and $SF_6$-$N_2$ arc plasmas are calculated based on our previous publications [20, 21]. Part of the data are illustrated in Figure 2. To demonstrate the generalization of Meta-PINN, we consider four cases corresponding to the arc models at different boundary temperatures, arc radii, arc pressures, and gas mixtures. In all cases, 200 and 500 scattered points are sampled uniformly along the radial coordinate as the support and query data, respectively. The number of inner iterations is fixed as 5 while the number of outer iterations is dependent on the cases we study. The learning rates of inner and outer optimization are set to be $10^{-5}$ and $10^{-4}$ respectively. For comparison with Meta-PINN, the normal PINN-based networks are initialized by the Xavier initialization. The random seed is set as a constant value for all the cases. The uncertainty arising from random seed should be investigated in the furture work. All the algorithms are implemented in the deep learning framework Pytorch [22]. Also for convenience of comparsion, we still perform a traditional calculation by FVM for all the cases. The relative $L^2$ error which is the relative error in $L^2$ norm between the PINN or Meta-PINN prediction and the solution by FVM, is also calculated for comparison in all cases.

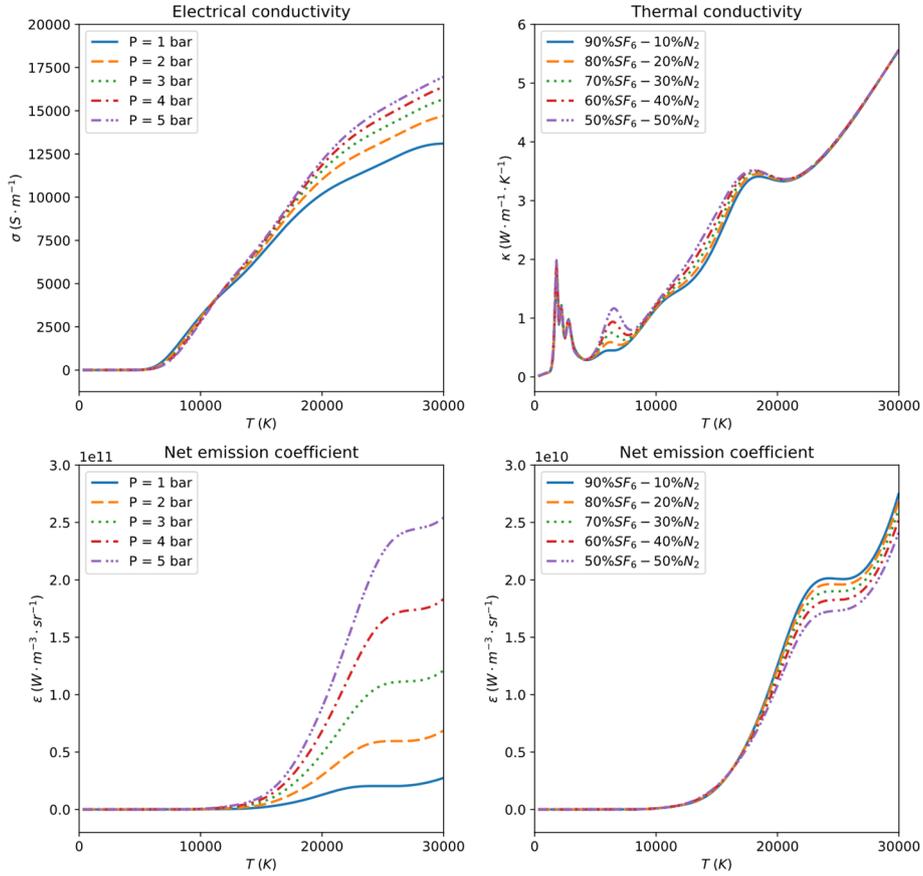

**Figure 2.** Transport and radiation coefficients of $SF_6$ and $SF_6$-$N_2$ arc plasmas at various conditions.

## 3.2. Solving arc model at different boundary temperatures

In this case, the training tasks are consisted of 1-D arc models at boundary temperatures $T_b$ = 1600 K, 1700 K, 1800 K, 1900 K, and 2000 K. The different boundary conditions result in different distributions of arc temperatures. In normal PINNs, the neural network trained with the boundary condition A cannot be used to directly predict the results with the new boundary condition B. To address this issue, we train a meta network on the training tasks for various boundary temperatures $T_b$, and the well-trained meta network is then used to yield good initial weights for PINN-based arc model with new $T_b$. This well-trained meta network is achieved after 20,000 epochs of outer optimization and tested at $T_b$ = 1550 K, 1650 K, 1750 K, 1850 K, 1950 K, and 2050 K. As we can see from the evolution of loss value and relative $L^2$ error in Figure 3, the training of PINNs for arc model with different boundary temperatures is improved significantly, not only in efficiency but also in accuracy. Table 1 in the appendix quantatively compares the optimizing epochs to reach a given loss value and relative $L^2$ error for PINN and Meta-PINN. It is verified that the neural network initialized by Meta-PINN converges faster than PINN even for the model at the boundary temperatures (e.g. $T_b$ = 1550 K and 2050 K) outside of training range. In terms of relative $L^2$ error ($10^{-2}$ and $5\times10^{-3}$), the minimum and maximum speed-up reach 1.7× and 6.9× respectively. It is worth mentioning that Meta-PINN can lower the relative $L^2$ error to $10^{-3}$ at most $T_b$ while PINN does not. It is also found that due to the faster convergence than normal PINNs, Meta-PINN presents more significant oscillation of loss and relative $L^2$ error during the training. This phenomena also exist in the other three cases.

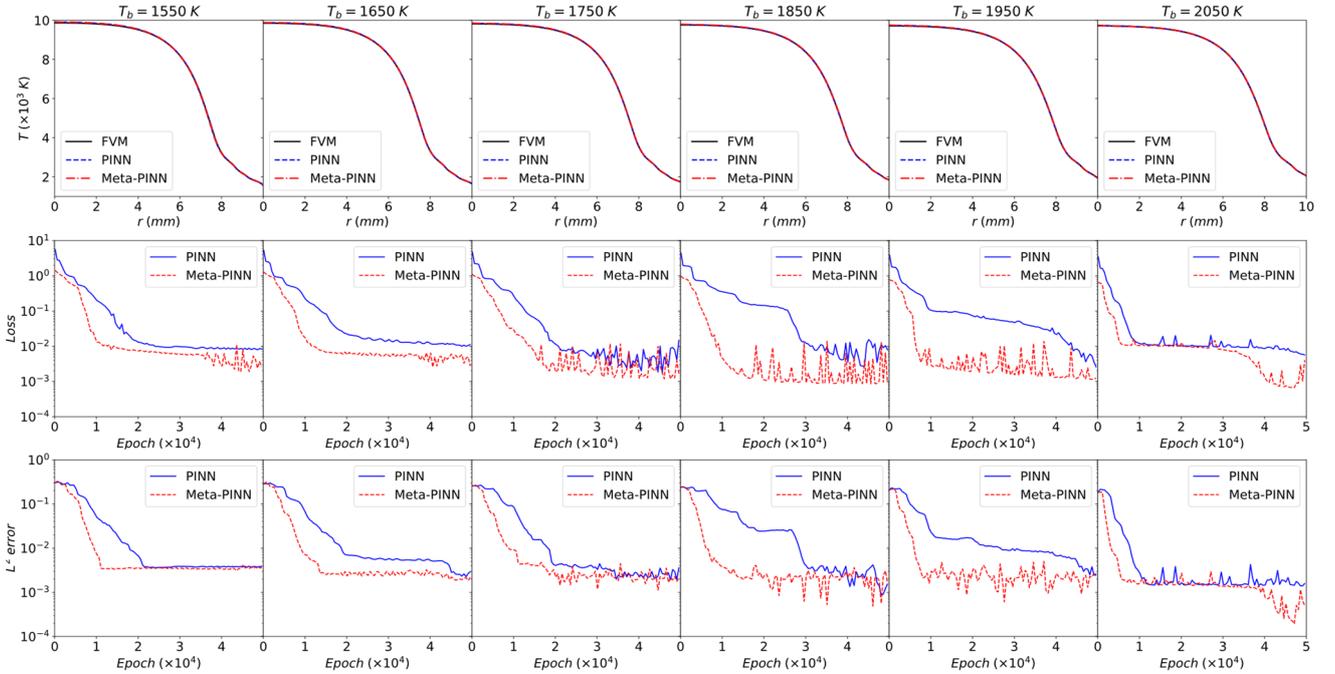

**Figure 3.** Test results of 1-D arc model at different boundary temperatures ($T_b$ = 1550 K, 1650 K, 1750 K, 1850 K, 1950 K, and 2050 K). First row: arc temperature predicted by FVM, PINN, and Meta-PINN. Second row:

loss value evolution with epochs for PINN and Meta-PINN. Third row: relative $L^2$ error evolution with epochs for PINN and Meta-PINN.

### 3.3. Solving arc model at different arc radii

In this case, the training tasks are composed of 1-D arc models at arc radii $R$ = 11 mm, 12 mm, 13 mm, 14 mm, and 15 mm. The arc radius influence not only the computational domain but also the net emission coefficients which are calculated at a given arc radius. As a result, the PINNs trained at arc radius A cannot infer the model results at arc radius B. In the framework of Meta-PINN, we train a meta network with the training tasks for various arc radii $R$, and then use it to initialize the PINN-based arc model with new $R$. The training of the meta network needs 30,000 epochs of outer optimization in our work. We test the meta model at $R$ = 10.5 mm, 11.5 mm, 12.5 mm, 13.5 mm, 14.5 mm, and 15.5 mm. As shown in Figure 4, although the unstable convergence is observed for both PINN and Meta-PINN, the training time of Meta-PINN is still reduced remarkably even for the model at the arc radii (e.g. $R$ = 10.5 mm and 15.5 mm) outside of training range. It is found in Table 2 in the appendix that the minimum and maximum speed-up in terms of relative $L^2$ error can reach 1.2× and 2.0× respectively.

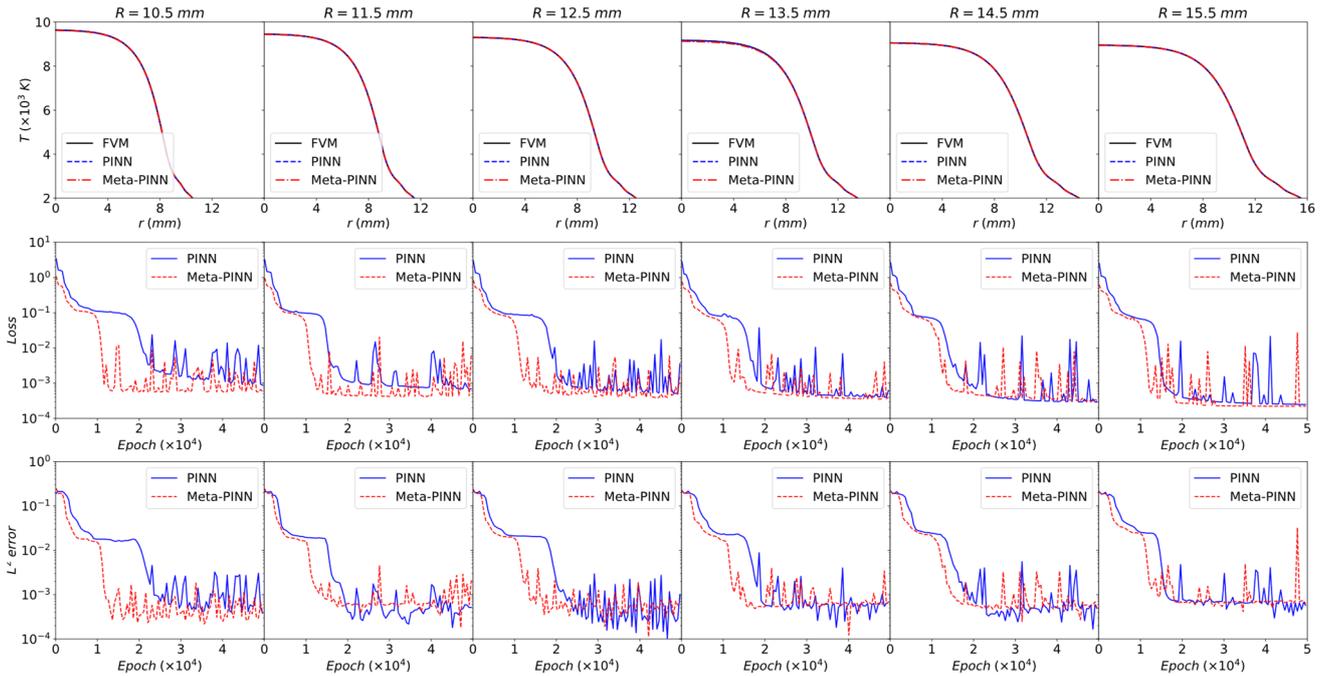

**Figure 4**. Test results of 1-D arc model at different arc radii ($R$ = 10.5 mm, 11.5 mm, 12.5 mm, 13.5 mm, 14.5 mm, and 15.5 mm). First row: arc temperature predicted by FVM, PINN, and Meta-PINN. Second row: loss value evolution with epochs for PINN and Meta-PINN. Third row: relative $L^2$ error evolution with epochs for PINN and Meta-PINN.

### 3.4. Solving arc model at different pressures

In this case, we consider the training tasks of 1-D arc models at arc pressures $P$ = 1 bar, 2 bar, 3 bar, 4 bar,

and 5 bar. As observed in Figure 2, the plasma properties, e.g. electrical conductivity and net emission coefficients, are strongly dependent on pressures. This means that the variation of arc pressures will lead to large variation of arc temperatures. The model trained at pressure A is not applicable at pressure B. Therefore, we perform a meta-learning on the training tasks and produce a meta network to initialize the weights for testing tasks at $P$ = 0.5 bar, 1.5 bar, 2.5 bar, 3.5 bar, 4.5 bar, and 5.5 bar. The meta network is trained by 20,000 epochs of outer optimization. Figure 5 describes the predicted arc temperatures, loss values, and relative $L^2$ error. Obviously, Meta-PINN can improve the training of PINNs extraordinarily both in efficiency and accuracy. With the same training consumption, Meta-PINN can yield better results than the normal PINN. For example, the relative $L^2$ error of the arc temperatures predicted by PINN at 2.5 bar is close to 0.9 after 50,000 epochs of training, whereas the error by Meta-PINN can be reduced to about $5\times10^{-4}$. In terms of relative $L^2$ error, the minimum and maximum speed-up can reach 1.2× and 2.7× respectively, except for the model at 5.5 bar. This exception could be attributed to the testing arc pressure outside of training range.

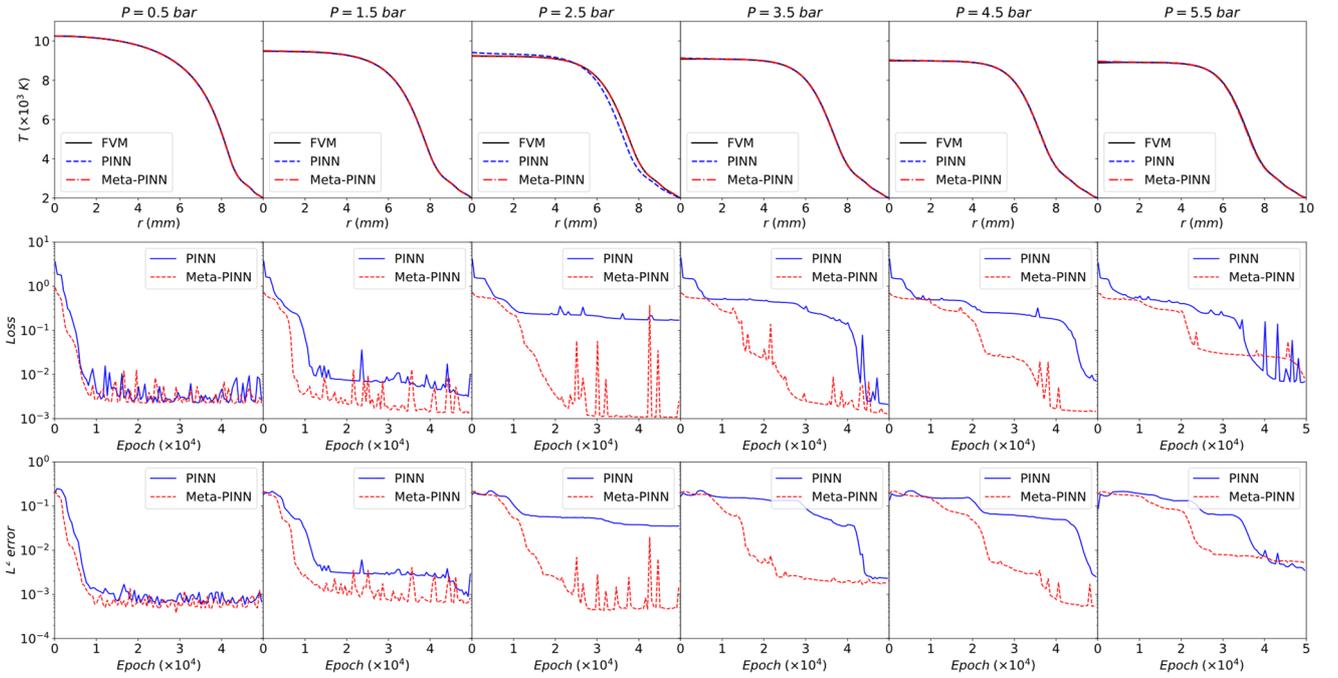

**Figure 5**. Test results of 1-D arc model at different pressures ($P$ = 0.5 bar, 1.5 bar, 2.5 bar, 3.5 bar, 4.5 bar, and 5.5 bar). First row: arc temperature predicted by FVM, PINN, and Meta-PINN. Second row: loss value evolution with epochs for PINN and Meta-PINN. Third row: relative $L^2$ error evolution with epochs for PINN and Meta-PINN.

### 3.5. Solving arc model at different gas mixtures

In the last case, we consider the training tasks with different gas mixtures, i.e. $SF_6$-$N_2$ arc plasmas. The mixing ratios of $SF_6$-$N_2$ are 90:10, 80:20, 70:30, 60:40, and 50:50 in the training tasks. As illustrated in Figure 2, the plasma properties vary significantly with the mixing ratios. To improve the training in this case, we also first train a meta network on the training tasks for various gas mixtures. The acceptable model is achieved after 100,000 epochs of outer optimization. We then use it to give initial weights for PINN-based models of $SF_6$-$N_2$ arc with mixing ratios of 95:5, 85:15, 75:25, 65:35, 55:45, and 45:55. The corresponding results are presented in Figure 6 and Table 4 in the appendix, and the remarkable improvement by Meta-PINN can be observed even for the model with mixing ratios outside of training range. Quantatively, the minimum and maximum speed-up in terms of relative $L^2$ error can reach 1.1× and 2.8× respectively. The greatest improvement is observed with the arc plasma of 55%$SF_6$-45%$N_2$. After 50,000 epochs of training, the relative $L^2$ error of this model predicted by PINN is approximately 0.4, whereas the error by Meta-PINN is reduced to about $3\times10^{-4}$.

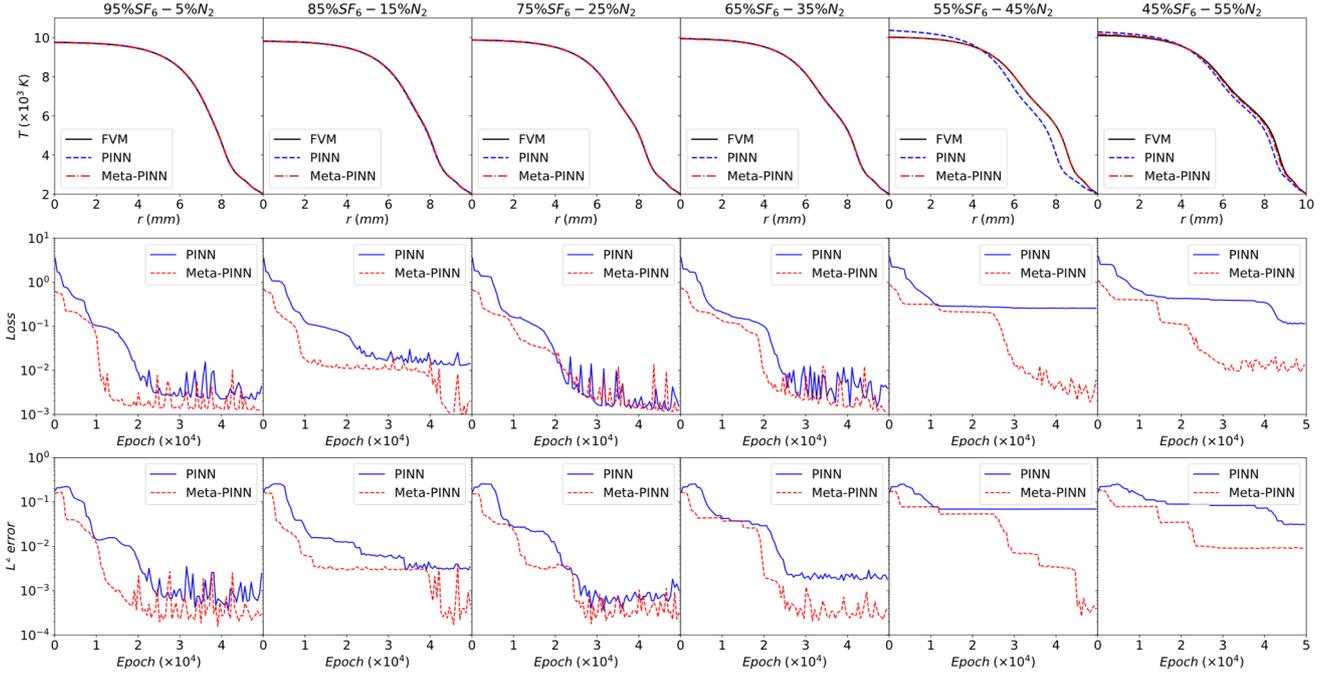

**Figure 6**. Test results of 1-D arc model at different gas mixtures (95%$SF_6$-5%$N_2$, 85%$SF_6$-15%$N_2$, 75%$SF_6$-25%$N_2$, 65%$SF_6$-35%$N_2$, 55%$SF_6$-45%$N_2$, and 45%$SF_6$-55%$N_2$). First row: arc temperature predicted by FVM, PINN, and Meta-PINN. Second row: loss value evolution with epochs for PINN and Meta-PINN. Third row: relative $L^2$ error evolution with epochs for PINN and Meta-PINN.

## 4. Conclusions

In this work, we propose a meta-learning method, namely Meta-PINN, for accelerating PINN-based plasma simulation. Meta-PINN first trains a meta network by a two-loop optimization on various training tasks. For each task, the training dataset is divided into support data and query data. The inner loop is performed on support data for the PINN optimization, and the outer loop is performed on query data for the meta optimization. A well-trained meta network is then used to initialize the PINN-based network for new tasks. We verify the ability of Meta-PINN by four cases of solving 1-D arc models at different boundary temperatures, arc radii, arc pressures, and gas mixtures. It is observed that a well-trained meta network can yield good initial weights for PINN-based 1-D arc models even at conditions slightly outside of training range. The acceleration performance depends on the difficulty of the tasks. The speed-up in terms of relative $L^2$ error by Meta-PINN ranges from 1.1× to 6.9× in the cases we studied. It should be noted that for the testing tasks with significant departure from training tasks, the generalization of Meta-PINN is still indeterminate, and should be further investigated in the furture work.

**Acknowledgments**

This work was supported in part by the National Natural Science Foundation of China (92066106, 51907023), the Young Scientific and Technical Talents Promotion Project of Jiangsu Association for Science and Technology (2021031), the Zhishan Young Scholar Project of Southeast University (2242022R40022), and the Fundamental Research Funds for the Central Universities (2242022R40022).

**Appendix A. Comparison between PINN and Meta-PINN**

**Table 1.** Optimizing epochs to reach given loss value and relative $L^2$ error in 1-D arc model at different boundary Temperatures $T_b$: Comparison between PINN and Meta-PINN

| | $T_b$ (K) | | 1550 | 1650 | 1750 | 1850 | 1950 | 2050 |
|---|---|---|---|---|---|---|---|---|
| Loss | $10^{-1}$ | PINN | 1.33 | 1.39 | 1.28 | 2.57 | 1.11 | 0.58 |
| | | Meta-PINN | 0.74 | 0.75 | 0.66 | 0.52 | 0.49 | 0.24 |
| | $10^{-2}$ | PINN | 2.41 | 4.78 | 2.04 | 3.11 | 4.39 | 2.50 |
| | | Meta-PINN | 1.13 | 1.21 | 1.39 | 0.86 | 0.63 | 1.68 |
| | $5\times10^{-3}$ | PINN | / | / | 2.89 | 3.75 | 4.61 | / |
| | | Meta-PINN | 3.42 | 3.17 | 1.55 | 0.99 | 0.68 | 3.58 |
| $L^2$ error | $10^{-2}$ | PINN | 1.79 | 1.87 | 1.83 | 2.86 | 2.68 | 0.77 |
| | | Meta-PINN | 0.92 | 0.90 | 0.88 | 0.60 | 0.61 | 0.33 |
| | $5\times10^{-3}$ | PINN | 2.08 | 4.15 | 1.91 | 2.96 | 4.51 | 0.86 |
| | | Meta-PINN | 1.08 | 1.22 | 1.11 | 0.76 | 0.65 | 0.47 |
| | $10^{-3}$ | PINN | / | / | / | 4.81 | / | / |
| | | Meta-PINN | / | / | 3.35 | 1.39 | 1.48 | 3.88 |

**Table 2.** Optimizing epochs to reach given loss value and relative $L^2$ error in 1-D arc model at different arc radii $R$: Comparison between PINN and Meta-PINN

| | $R$ (mm) | | 10.5 | 11.5 | 12.5 | 13.5 | 14.5 | 15.5 |
|---|---|---|---|---|---|---|---|---|
| Loss | $10^{-1}$ | PINN | 1.44 | 0.87 | 0.71 | 0.63 | 0.55 | 0.58 |
| | | Meta-PINN | 0.84 | 0.57 | 0.51 | 0.49 | 0.45 | 0.41 |
| | $10^{-2}$ | PINN | 2.07 | 1.53 | 1.86 | 1.61 | 1.40 | 1.44 |
| | | Meta-PINN | 1.09 | 1.07 | 1.10 | 1.12 | 1.17 | 1.13 |
| | $5\times10^{-3}$ | PINN | 2.17 | 1.56 | 1.89 | 1.66 | 1.44 | 1.46 |
| | | Meta-PINN | 1.12 | 1.09 | 1.14 | 1.13 | 1.21 | 1.14 |

|  |  |  | | | | | |
|---|---|---|---|---|---|---|---|
| L² error | 10⁻² | PINN | 2.05 | 1.53 | 1.86 | 1.62 | 1.42 | 1.46 |
| | | Meta-PINN | 1.09 | 1.06 | 1.11 | 1.13 | 1.19 | 1.19 |
| | 5×10⁻³ | PINN | 2.13 | 1.57 | 1.92 | 1.69 | 1.51 | 1.51 |
| | | Meta-PINN | 1.12 | 1.10 | 1.16 | 1.15 | 1.25 | 1.25 |
| | 10⁻³ | PINN | 2.36 | 1.90 | 2.13 | 1.88 | 1.82 | 1.82 |
| | | Meta-PINN | 1.21 | 1.37 | 1.34 | 1.29 | 1.51 | 1.51 |

**Table 3.** Optimizing epochs to reach given loss value and relative $L^2$ error in 1-D arc model at different pressures $P$: Comparison between PINN and Meta-PINN

| | $P$ (bar) | | 0.5 | 1.5 | 2.5 | 3.5 | 4.5 | 5.5 |
|---|---|---|---|---|---|---|---|---|
| Loss | 10⁻¹ | PINN | 0.50 | 0.97 | / | 4.09 | 4.33 | 3.43 |
| | | Meta-PINN | 0.27 | 0.65 | 1.20 | 1.47 | 2.13 | 2.17 |
| | 10⁻² | PINN | 0.69 | 1.21 | / | 4.28 | 4.71 | 4.07 |
| | | Meta-PINN | 0.66 | 0.72 | 1.69 | 2.39 | 3.46 | 4.92 |
| | 5×10⁻³ | PINN | 0.88 | 4.01 | / | 4.35 | / | / |
| | | Meta-PINN | 0.74 | 0.83 | 1.88 | 2.58 | 3.58 | / |
| L² error | 10⁻² | PINN | 0.61 | 1.15 | / | 4.30 | 4.66 | 3.81 |
| | | Meta-PINN | 0.46 | 0.72 | 1.37 | 1.59 | 2.34 | 2.59 |
| | 5×10⁻³ | PINN | 0.65 | 1.27 | / | 4.37 | 4.75 | 4.24 |
| | | Meta-PINN | 0.55 | 0.78 | 1.61 | 1.99 | 2.58 | 4.99 |
| | 10⁻³ | PINN | 1.15 | 4.81 | / | / | / | / |
| | | Meta-PINN | 0.84 | 1.94 | 2.37 | / | 3.85 | / |

**Table 4.** Optimizing epochs to reach given loss value and relative $L^2$ error in 1-D arc model at different proportions of $N_2$: Comparison between PINN and Meta-PINN

| | Proportion of $N_2$ (%) | | 5 | 15 | 25 | 35 | 45 | 55 |
|---|---|---|---|---|---|---|---|---|
| Loss | 10⁻¹ | PINN | 1.1 | 1.38 | 1.57 | 2.01 | / | / |
| | | Meta-PINN | 0.91 | 0.75 | 0.97 | 1.48 | 2.66 | 2.20 |
| | 10⁻² | PINN | 1.93 | / | 2.16 | 2.41 | / | / |
| | | Meta-PINN | 1.08 | 3.53 | 2.23 | 2.00 | 3.12 | 3.15 |
| | 5×10⁻³ | PINN | 2.07 | / | 2.29 | 2.54 | / | / |
| | | Meta-PINN | 1.13 | 4.06 | 2.43 | 2.13 | 3.62 | / |
| L² error | 10⁻² | PINN | 1.81 | 2.20 | 2.02 | 2.29 | / | / |
| | | Meta-PINN | 1.05 | 0.87 | 1.13 | 1.93 | 2.86 | 2.43 |
| | 5×10⁻³ | PINN | 1.98 | 3.39 | 2.14 | 2.47 | / | / |
| | | Meta-PINN | 1.09 | 1.21 | 1.30 | 1.96 | 3.59 | / |

|  |  | PINN | 2.33 | / | 2.61 | / | / | / |
|  | $10^{-3}$ | Meta-PINN | 1.46 | 4.03 | 2.43 | 2.41 | 4.51 | / |

## REFERENCES

[1]	Derouillat J, Beck A, Pérez F, Vinci T, Chiaramello M, Grassi A, Flé M, Bouchard G, Plotnikov I and Aunai N 2018 Smilei: A collaborative, open-source, multi-purpose particle-in-cell code for plasma simulation *Computer Physics Communications* **222** 351-73

[2]	Gleizes A, Gonzalez J J and Freton P 2005 Thermal plasma modelling *Journal of Physics D: Applied Physics* **38** R153-R83

[3]	Raissi M, Perdikaris P and Karniadakis G E 2019 Physics-informed neural networks: A deep learning framework for solving forward and inverse problems involving nonlinear partial differential equations *Journal of Computational Physics* **378** 686-707

[4]	Raissi M, Yazdani A and Karniadakis G E 2020 Hidden fluid mechanics: Learning velocity and pressure fields from flow visualizations *Science* **367** 1026

[5]	Pun G P P, Batra R, Ramprasad R and Mishin Y 2019 Physically informed artificial neural networks for atomistic modeling of materials *Nature Communications* **10** 2339

[6]	De Florio M, Schiassi E, Ganapol B D and Furfaro R 2021 Physics-informed neural networks for rarefied-gas dynamics: Thermal creep flow in the Bhatnagar–Gross–Krook approximation *Physics of Fluids* **33** 047110

[7]	Arzani A, Wang J-X and D'Souza R M 2021 Uncovering near-wall blood flow from sparse data with physics-informed neural networks *Physics of Fluids* **33** 071905

[8]	Ji W, Qiu W, Shi Z, Pan S and Deng S 2021 Stiff-pinn: Physics-informed neural network for stiff chemical kinetics *The Journal of Physical Chemistry A* **125** 8098-106

[9]	Kawaguchi S, Takahashi K, Ohkama H and Satoh K 2020 Deep learning for solving the Boltzmann equation of electrons in weakly ionized plasma *Plasma Sources Science and Technology* **29** 025021

[10]	Zhang Y, Fu H, Qin Y, Wang K and Ma J 2022 Physics-Informed Deep Neural Network for Inhomogeneous Magnetized Plasma Parameter Inversion *IEEE Antennas and Wireless Propagation Letters* **21** 828-32

[11]	Zhong L, Wu B and Wang Y 2022 Low-temperature plasma simulation based on physics-informed neural networks: frameworks and preliminary applications *Physics of Fluids* **34** 087116

[12]	Zhong L, Gu Q and Wu B 2020 Deep learning for thermal plasma simulation: Solving 1-D arc model as an example *Computer Physics Communications* **257** 107496

[13]	Huisman M, van Rijn J N and Plaat A 2021 A survey of deep meta-learning *Artificial Intelligence Review*

[14]	Psaros A F, Kawaguchi K and Karniadakis G E 2022 Meta-learning PINN loss functions *Journal of Computational Physics* **458** 111121

[15]	Finn C, Abbeel P and Levine S 2017 Model-agnostic meta-learning for fast adaptation of deep networks. In: *International Conference on Machine Learning*: PMLR) pp 1126-35

[16]	Ni R, Goldblum M, Sharaf A, Kong K and Goldstein T 2021 Data augmentation for meta-learning. In: *International Conference on Machine Learning*: PMLR) pp 8152-61

[17]	Glorot X and Bengio Y 2010 Understanding the difficulty of training deep feedforward neural networks. In: *Proceedings of the thirteenth international conference on artificial intelligence and statistics,* pp 249-56

[18]	Li Z, Kovachki N, Azizzadenesheli K, Liu B, Bhattacharya K, Stuart A and Anandkumar A 2020 Fourier neural

|      | operator for parametric partial differential equations *arXiv:2010.08895* |
|------|---|
| [19] | Lu L, Jin P, Pang G, Zhang Z and Karniadakis G E 2021 Learning nonlinear operators via DeepONet based on the universal approximation theorem of operators *Nature Machine Intelligence* **3** 218-29 |
| [20] | Zhong L, Wang J, Xu J, Wang X and Rong M 2019 Effects of Buffer Gases on Plasma Properties and Arc Decaying Characteristics of C4F7N-N2 and C4F7N-CO2 Arc Plasmas *Plasma Chemistry and Plasma Processing* **39** 1379-96 |
| [21] | Zhong L, Cressault Y and Teulet P 2018 Thermophysical and radiation properties of high-temperature C4F8-CO2 mixtures to replace SF6 in high-voltage circuit breakers *Phys. Plasmas* **25** 033502 |
| [22] | Paszke A, Gross S, Massa F, Lerer A, Bradbury J, Chanan G, Killeen T, Lin Z, Gimelshein N and Antiga L 2019 PyTorch: An imperative style, high-performance deep learning library. In: *Advances in Neural Information Processing Systems,* (Vancouver, Canada pp 8024-35 |